\documentclass[preprint,showpacs,superscriptaddress,amsmath,amssymb]{revtex4}
\usepackage{graphicx}
\usepackage{bm}

\begin{document}

\title{On soliton structure of the Vakhnenko equation with a 'dissipative' term: a peculiar fission phenomenon}

\author{Kuetche Kamgang Victor}
\email{vkuetche@yahoo.fr}

\affiliation{Department of Physics, Faculty of Science, University
of Yaounde I, P.O. Box. 812, Cameroon}

\author{Bouetou Bouetou Thomas}
\email{tbouetou@yahoo.fr}

\affiliation{Ecole Nationale Sup$\acute{e}$rieure Polytechnique,
University of Yaounde I, P.O. Box. 8390, Cameroon}

\affiliation{The Abdus Salam International Centre for Theoretical
Physics, P.O. Box 586, Strada Costiera, II-34014, Trieste, Italy}

\author{Timoleon Crepin Kofane}
\email{tckofane@yahoo.com}

\affiliation{Department of Physics, Faculty of Science, University
of Yaounde I, P.O. Box. 812, Cameroon}

\affiliation{The Abdus Salam International Centre for Theoretical
Physics, P.O. Box 586, Strada Costiera, II-34014, Trieste, Italy}

\date{\today}

\begin{abstract}
In the present work, we investigate the soliton structure of the
Vakhnenko equation with a 'dissipative' term, by means of the
Hirota's method. As a result, we unearth three kinds of soliton
solutions depending upon the dissipation parameter. We further
find that the scattering behavior among such structures exhibits a
peculiar feature that may be the fission phenomenon.
\end{abstract}

\pacs{05.45.Yv}

\maketitle

\section{\label{sec1}INTRODUCTION}
Nonlinear systems and their related waves in different branches of
physics have recently been investigated in detail through theory
and experiments. Since the investigation of the nonlinear
transmission line, pioneered by Hirota and Sazuki \cite{HSU},
various, kinds of nonlinear wave propagation have been discussed
\cite{NEJ,YNO,BPA,YKA}. Integrable models of such systems have
been applied to many fields of research such as condensed matter
\cite{LRO,COF}, fluid mechanics \cite{TMA}, plasma physics
\cite{DAS}, optics \cite{GSB,GEO}, communications \cite{NKO,DFW},
chemistry \cite{KKP}, and biology \cite{KHK,MUT}. These integrable
models which may take the form of nonlinear partial differential
equations, may possess some kinds of self-confined solitary waves
solutions known as solitons.

Since Gardner, Greene, Kruskal and Miura have solved the
Korteweg-de Vries equation using the inverse scattering method
\cite{GGK}, the modern theory of solitons has been deeply studied
in mathematics, and widely applied in physics. A wealth of
interesting properties of soliton equations have been found .
Indeed, a soliton equation may possess B$\ddot{a}$cklund and
Darboux transformations, the Lax-pair, N-soliton solutions, a
bilinear form, a zero-curvature form, the Painlev$\acute{e}$
property, infinitely many symmetries and conservation laws, just
to name a few \cite{ACL}.

Recently, Vakhnenko \cite{VAK1} have investigated high-frequency
perturbations in a relaxing barothropic medium. He has then
derived a new nonlinear evolution equation known as Vakhnenko
equation \cite{VAK1} given by
\begin{eqnarray}
u_{tx}+\left(uu_{x}\right)_{x}+\alpha u_{x}+u=0,\label{eq0.1}
\end{eqnarray}
where $u$ may stand for a physical observable, and subscripts $t$
and $x$ appended to $u$ may refer to partial differentiation with
respect to these independent variables. It may be noted here that
the $\alpha u_{x}$-term may be regarded and defined as the
'dissipation' term \cite{VAK1}. Without 'dissipation', this
equation which has been subject to many investigations
\cite{PAR,VAK2,VP1,VP2,VP3}, may possess loop-like soliton
solutions. As far as we are concerned, few investigations
\cite{VAK1} of Eq. (\ref{eq0.1}) may have been done. Vakhnenko
\cite{VAK1} has pointed out that Eq. (\ref{eq0.1}) may possess
ambiguous solutions in the form of a solitary wave, and he has
proved that the dissipative term, with a dissipative parameter
$\alpha$ less than that limit value, does not destroy the
loop-like solutions. One underlying query that may obviously be
asked may be related to the existence of other kinds of solutions
when the dissipation parameter is greater or equal to this limit
value. Besides, the question of soliton structure of such
solutions may be investigated.

In order to provide some answers to these queries, the study of
the fundamental role played by the dissipation parameter may be
done around the following points. In Sec. \ref{sec2}, the
different kinds of the one-soliton solution are presented. In Sec.
\ref{sec3}, the two-soliton solutions and their scattering
behavior are depicted. Some peculiar phenomenon such as the
fission phenomenon, has been identified.  Sec. \ref{sec4} is
devoted to a brief summary.

\section{\label{sec2}BILINEARIZATION OF THE VAKHNENKO EQUATION AND CONSTRUCTION OF THE ONE-SOLITON SOLUTIONS}
Introducing new independent variables $X$ and $T$ as follows
\cite{VP1}
\begin{eqnarray}
x=T+\int_{-\infty}^{X}U(s,T)ds+x_{0},\quad t=X,\label{eq1.1}
\end{eqnarray}
where $u(t,x)\equiv U(X,T)$, and $x_{0}$ stands for an arbitrary
constant, setting $U=W_{X}$, Eq. (\ref{eq0.1}) transformed to
\cite{VP1}
\begin{eqnarray}
W_{XXT}+W_{X}W_{T}+\alpha W_{T}+W_{X}=0.\label{eq1.2}
\end{eqnarray}
By taking
\begin{eqnarray}
W=6\left(\ln F\right)_{X},\label{eq1.3}
\end{eqnarray}
Eq. (\ref{eq1.2}) may take the following coupled bilinearized
forms
\begin{subequations}
\label{eq1.4}
\begin{equation}
\left(D_{T}D_{X}^{3}+D_{X}^{2}\right)F\cdot F+\alpha
GF=0,\label{eq1.4a}
\end{equation}
\begin{equation}
D_{T}\left(D^{2}_{X}F\cdot F\right)\cdot
F^{2}-GF^{3}=0,\label{eq1.4b}
\end{equation}
\end{subequations}
where $G$ may stand for an arbitrary function. The quantities
$D_{T}$ and $D_{X}$ denote the Hirota operators. According to the
usual procedure, the soliton solutions may be constructed by
expanding $F$ and $G$ in suitable formal power series.

Thus, the one-soliton solutions may be given by
\begin{eqnarray}
F= 1+\exp(2\eta),\label{eq1.5}
\end{eqnarray}
where
\begin{eqnarray}
\eta=KX-\omega T+\eta_{0},\label{eq1.6}
\end{eqnarray}
quantities $K$ and $\omega$ standing for wave number and angular
frequency, respectively. The dispersion equation may be given by
\begin{eqnarray}
2\left(\alpha+2K\right)\omega-1=0,\label{eq1.7}
\end{eqnarray}
and $\omega=Kv$, $v>0$ being the velocity of the wave. Combining
Eqs. (\ref{eq1.3}) and (\ref{eq1.5}) may lead to
\begin{eqnarray}
W=6K\left[1+\tanh\left(\eta\right)\right],\label{eq1.8}
\end{eqnarray}
and hence
\begin{eqnarray}
U=U_{M}\verb"sech"^{2}\left(\eta\right),\label{eq1.9}
\end{eqnarray}
where the amplitude $U_{M}=6K^{2}$.

In order to discuss the different types of solutions, it seems
worthy to consider the following equation
\begin{eqnarray}
\partial_{T}=\left[1-6K\omega\verb"sech"^{2}\left(\eta\right)\right]\partial_{x}.\label{eq1.10}
\end{eqnarray}
Thus, setting $\lambda=6K\omega$, we may distinguish three kinds
of solutions according to the cases $\lambda<1$, $\lambda=1$, and
$\lambda>1$. As a result, we find that
\begin{enumerate}
    \item for $\alpha<\frac{1}{\sqrt{6v}}$, loop-like solution may be
obtained (see FIG. 1 where loop is represented by the solid line);
    \item for $\alpha=\frac{1}{\sqrt{6v}}$,
cusp-like solution may be obtained (see FIG. 1 where cusp is
represented by the broken line);

\item finally, for $\alpha>\frac{1}{\sqrt{6v}}$, hump-like
solution may be obtained (see FIG. 1 where hump is represented by
the dotted line).
\end{enumerate}
In FIG. 1, the previous kinds of soliton solutions have been
depicted as variations of $U/U_{M}$ vs $x$. It seems also worth
depicting the domain in which these solutions may exist. In this
view, we depict at FIG. 2, the variations of $\alpha$ vs the
velocity of waves $v$. In this figure, the white area may refer to
hump-like soliton solutions. It seems also worthy to compare the
amplitudes of these solutions. The variations of $U_{M}$ vs the
velocity of waves $v$ are depicted in FIG. 3 where the white area
may refer to the loop-like soliton solutions. Thus, for a given
velocity, the greatest amplitude may be due to loop-like solutions
and the smallest amplitude belongs to hump-like solutions.

\section{\label{sec3}THE TWO-SOLITON SOLUTIONS}
The solution to Eq. (\ref{eq1.4}) corresponding to two-soliton
solutions may be given by
\begin{eqnarray}
F=1+\exp(\eta_{2})+\exp(2\eta_{2})+A\exp(4\eta_{1})+B\exp(4\eta_{2})+C\exp2(\eta_{1}+\eta_{2}),\label{eq2.1}
\end{eqnarray}
where the phases $\eta_{i}$ $(i=1, 2)$ may be given by
\begin{equation}
\eta_{i}=K_{i}X-\omega_{i}T+\eta_{i0},\quad (i=1,2),\label{eq2.2}
\end{equation}
and the dispersion relations
\begin{equation}
2\left(\alpha+2K_{i}\right)\omega_{i}-1=0,\quad
(i=1,2)\label{eq2.3}
\end{equation}
where $\omega_{i}=K_{i}v_{i}$ $(i=1, 2)$, $v_{i}$ standing for
velocities of waves.

The c$\oe$fficients $A$, $B$ and $C$ may be given by
\begin{subequations}
\label{eq2.4}
\begin{equation}
A=\frac{\alpha}{2\left(\alpha+6K_{1}\right)},\label{eq2.4a}
\end{equation}
\begin{equation}
B=\frac{\alpha}{2\left(\alpha+6K_{2}\right)},\label{eq2.4b}
\end{equation}
\begin{equation}
C=\frac{2\alpha\left[\left(\omega_{1}-\omega_{2}\right)\left(K_{1}^{2}-K_{2}^{2}\right)+2K_{1}K_{2}\left(\omega_{1}+\omega_{2}\right)\right]+4\left(\omega_{2}-\omega_{1}\right)\left(K_{1}-K_{2}\right)^{3}+\left(K_{1}-K_{2}\right)^{2}}{\left(K_{1}+K_{2}\right)^{2}\left[2\left(\omega_{1}+\omega_{2}\right)\left(\alpha+2K_{1}+K_{2}\right)-1\right]}.\label{eq2.4c}
\end{equation}
\end{subequations}

In order to investigate the scattering behavior among these
two-soliton solutions, we may firstly consider some underlying
snapshots.
\begin{itemize}
    \item For $\alpha=1.2$, according to Eq.
(\ref{eq1.10}), two snapshots depicting loop-like and hump-like
solitons may be depicted at FIGS. 4 and 5 at times $t=-15$ and
$t=11$, respectively;
    \item for $\alpha=0.1$, the same phenomenon may be observed but we may see that the loop-like soliton splits into two other loops of the same
    amplitudes. This case may be depicted at FIGS. 6 and 7 at times
$t=-15$ and $t=11$, respectively. Thus, as a result, there may
exist a value at which the loop-like soliton begins to split. It
may be suggested that for velocities $v_{1}=0.24$ and
$v_{2}=0.12$, there may not be splitting provided the relation
$1.2<\alpha<2.6$ holds, whereas for $\alpha<1.2$, splitting may be
observed;
    \item for $\alpha=2.6$ with velocities $v_{1}=0.24$ and
$v_{2}=0.12$, according to Eq. (\ref{eq1.10}), two snapshots
depicting cusp-like and hump-like solitons may be depicted at
FIGS. 8 and 9 at times $t=-15$ and $t=11$, respectively;
    \item finally for $\alpha=5.0$ with velocities $v_{1}=0.24$ and
$v_{2}=0.12$, according to Eq. (\ref{eq1.10}), two snapshots
depicting large hump-like and small hump-like solitons may be
depicted at FIGS. 10 and 11 at times $t=-15$ and $t=11$,
respectively.
\end{itemize}
Now, the full scattering behavior among these soliton solutions
may be properly analyzed. As it may be observed, for negative
values of time $t$, only one-soliton solution may be observed. For
positive values of time $t$, the initial one-soliton solution
splits into two other solitons with an increase in the amplitude.
The soliton with the small amplitude may have hump-like shape.
This kind of phenomenon that may refer to a fission, has recently
been investigated by Morrison and Parkes \cite{VP4} while deriving
the N-soliton solution to a generalized Vakhnenko equation.

\section{\label{sec4}SUMMARY}
In this paper, we have investigated the soliton structure of the
Vakhnenko equation \cite{VP1} with a 'dissipative' term, by means
of the Hirota's method. This model equation derived by Vakhnenko
\cite{VP1} may be underlying in description of high-frequency
perturbations in a relaxing barothropic medium. As a result, we
have unearthed three kinds of soliton solutions depending upon the
dissipation parameter. These soliton solutions may have loop-like,
cusp-like and hump-like shapes, respectively. We have further
found that the scattering behavior among such structures exhibits
a peculiar feature known as fission. Indeed, the initial
one-soliton splits into two other kinds of solitons as time
elapses from negative values to positive values. This kind of
phenomenon may have been pointed out by Morrison and Parkes
\cite{VP4} while investigating the N-soliton solution to a
generalized Vakhnenko equation. In order to provide the full
detail on the interactions such as asymptotic bahavior of the
scattering among the soliton solutions to the Vakhnenko equation
\cite{VP1} (see Eq. (\ref{eq0.1})), the same procedure described
in Ref. \cite{VP4} may be followed. Nonetheless, for some
convenience due to the length of the paper, we do not report such
result here. We may have only focused our interest to the effect
of the dissipative parameter $\alpha$ on the Vakhnenko equation
\cite{VP1} (see Eq. (\ref{eq0.1})).

For a further interest, it may be worthy to investigate the
N-soliton solutions $(N\geq 3)$ to the Vakhnenko equation
\cite{VP1} (see Eq. (\ref{eq0.1})). Another interesting study may
be to fix the dissipation parameter $\alpha$, and subsequently
investigate the effect of the velocities of waves on the system.
Besides, following a recent work of Konno and Kakuhata \cite{KKA}
on rotating soliton solutions to a coupled dispersionless system,
it may be worth considering complex-valued angular frequency
$\omega$ and wave number $K$ into the dispersion equation. This
may hopefully help investigating rotating soliton solutions to the
Vakhnenko equation \cite{VP1} (see Eq. (\ref{eq0.1})).

\bibliography{JMPVakhndisspM}
\newpage
\begin{figure}
    \begin{center}
    \includegraphics{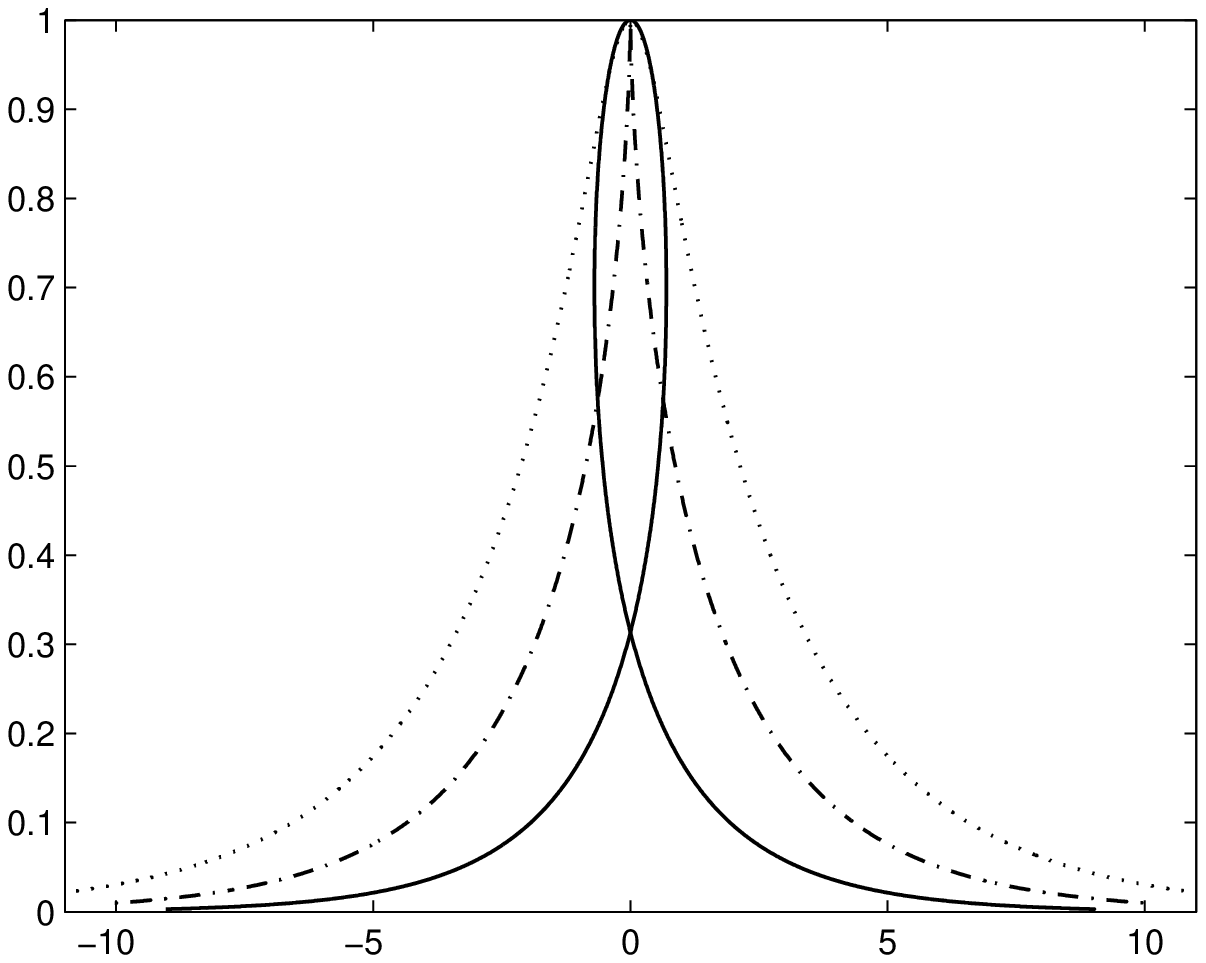}
    \caption{One-soliton solutions $U/U_{M}$ vs $x$: the solid, broken and dotted lines may represent the loop-like, cusp-like and hump-like soliton solutions, respectively.}\label{FIG1}
\end{center}
\end{figure}

\begin{figure}
    \begin{center}
    \includegraphics{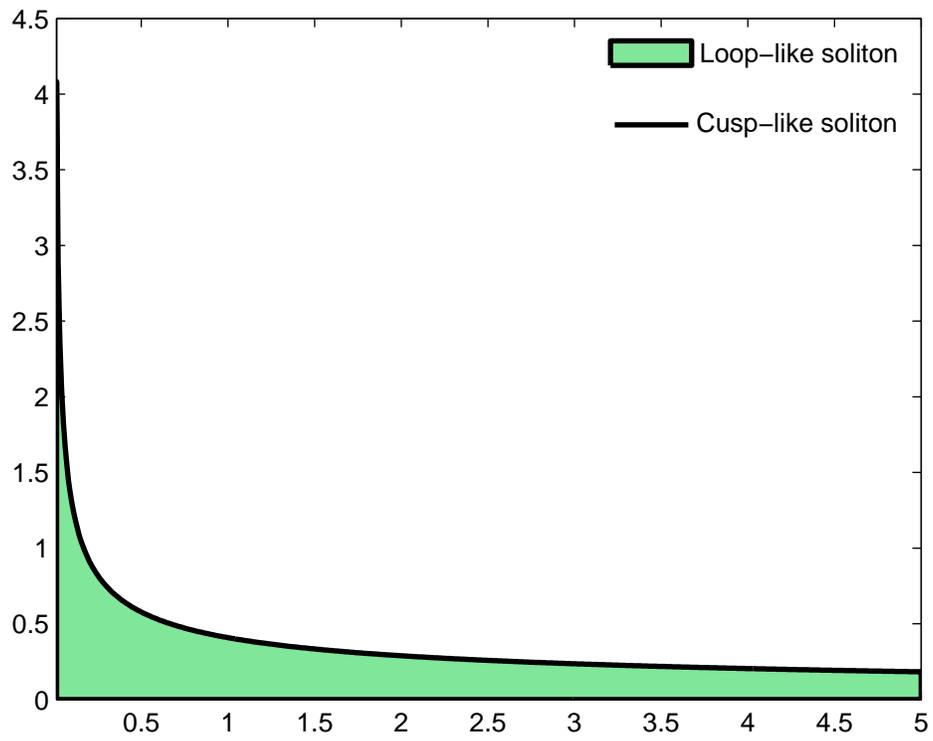}
    \caption{Variations of the dissipation parameter $\alpha$ vs the velocity $v$ of waves. White zone may refer to hump-like soliton solutions.}\label{FIG2}
\end{center}
\end{figure}

\begin{figure}
    \begin{center}
    \includegraphics{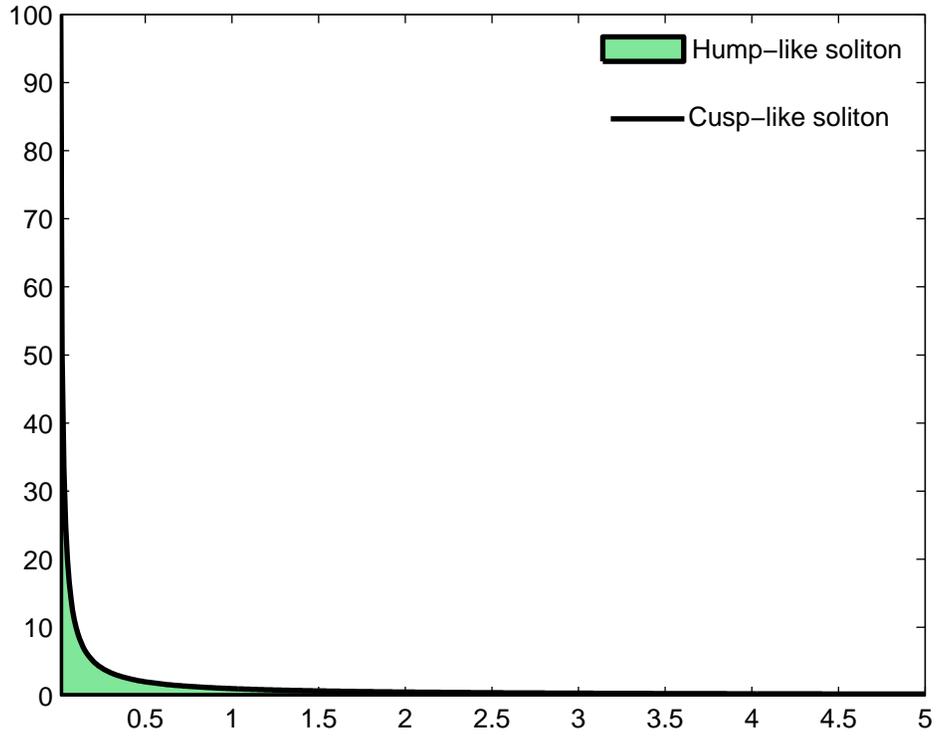}
    \caption{Variations of the amplitude $U_{M}$ of waves vs the velocity $v$ of waves. White area may refer to loop-like soliton solutions.}\label{FIG3}
\end{center}
\end{figure}

\begin{figure}
    \begin{center}
    \includegraphics{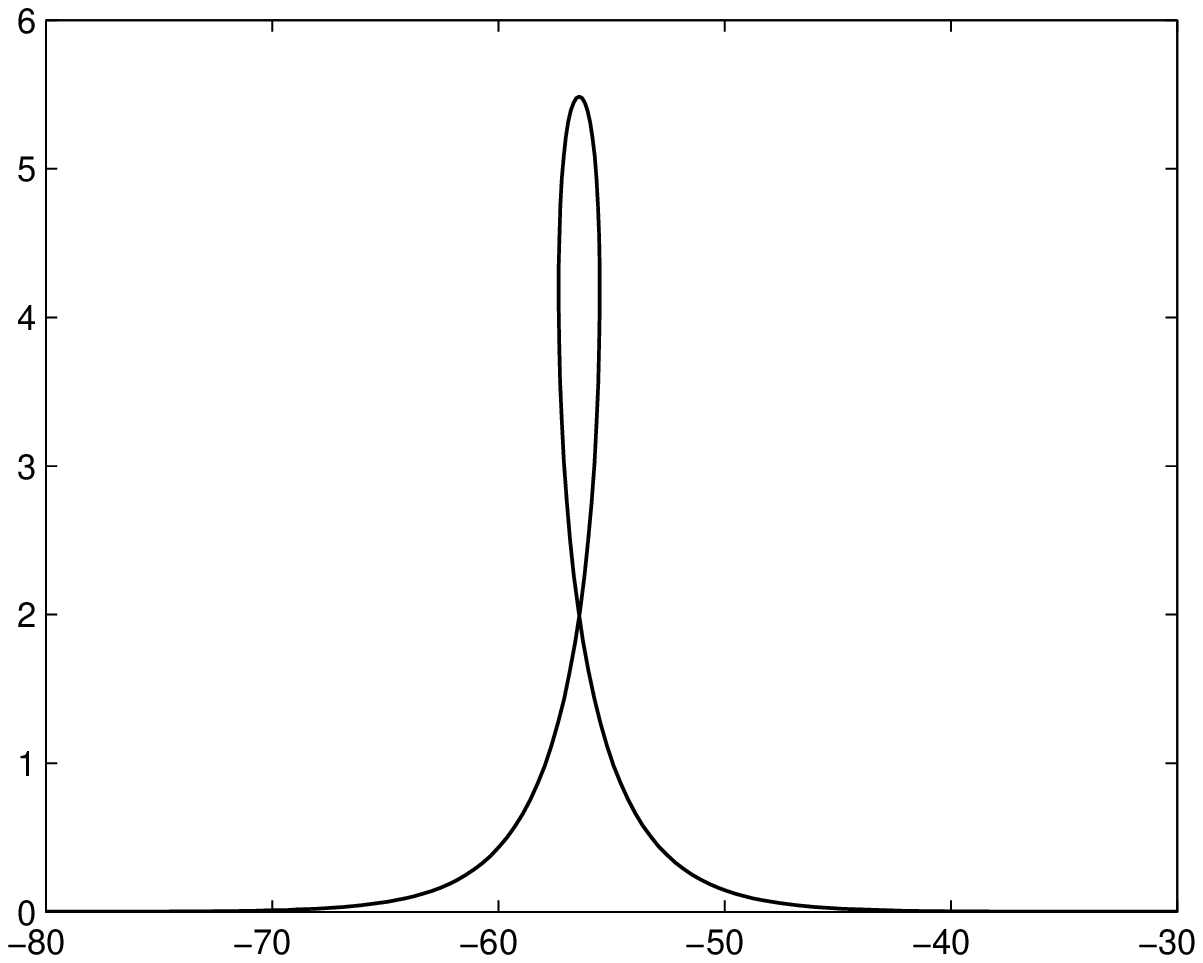}
    \caption{Snapshot of loop-like soliton solution at time $t=-15$ from Eq. (\ref{eq2.1}) for $\alpha=1.2$, $v_{1}=0.24$ and
$v_{2}=0.12$.}\label{FIG4}
\end{center}
\end{figure}

\begin{figure}
    \begin{center}
    \includegraphics{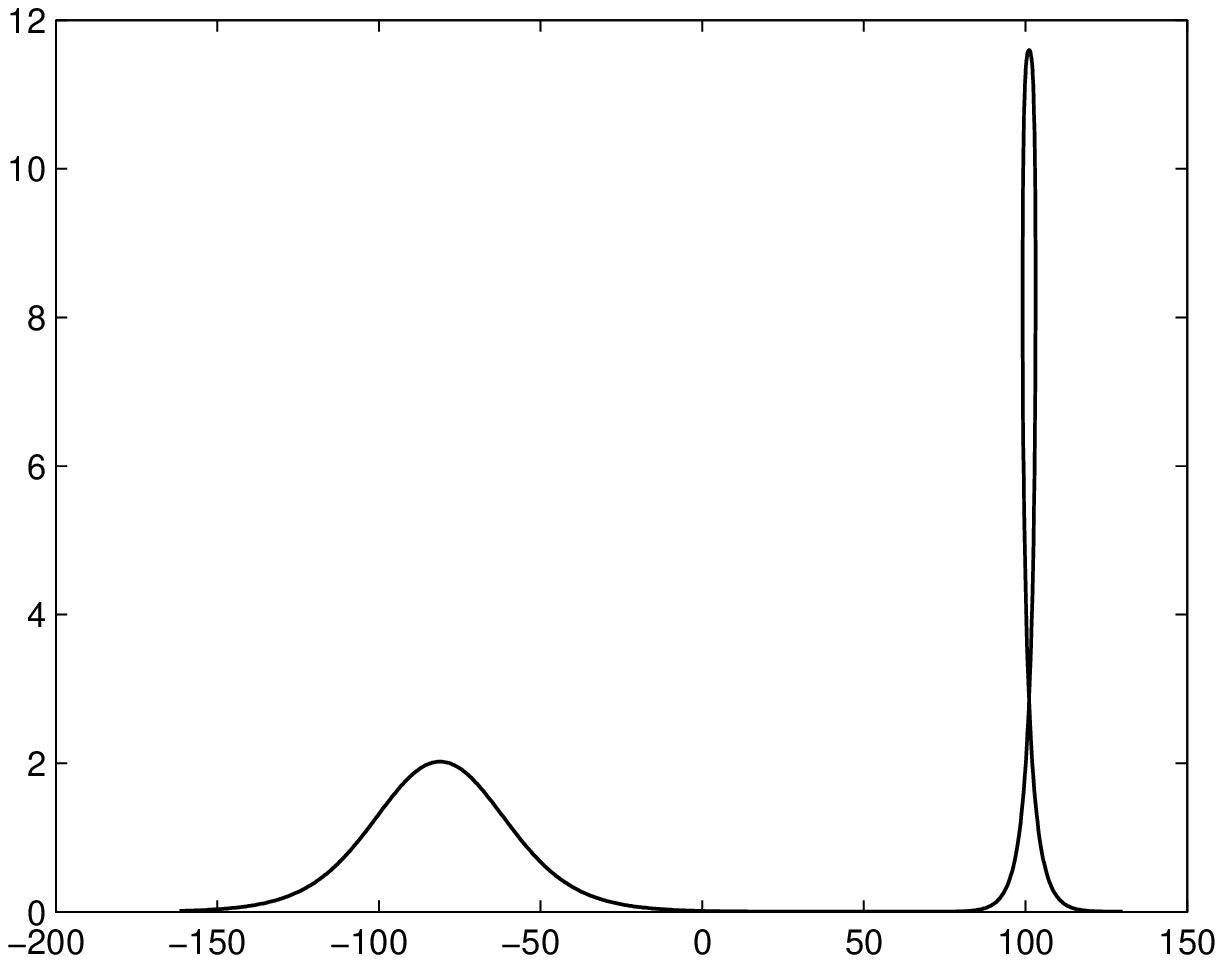}
    \caption{Snapshot of loop-like soliton solution at time $t=11$ from Eq. (\ref{eq2.1}) for $\alpha=1.2$, $v_{1}=0.24$ and
$v_{2}=0.12$.}\label{FIG5}
\end{center}
\end{figure}

\begin{figure}
    \begin{center}
    \includegraphics{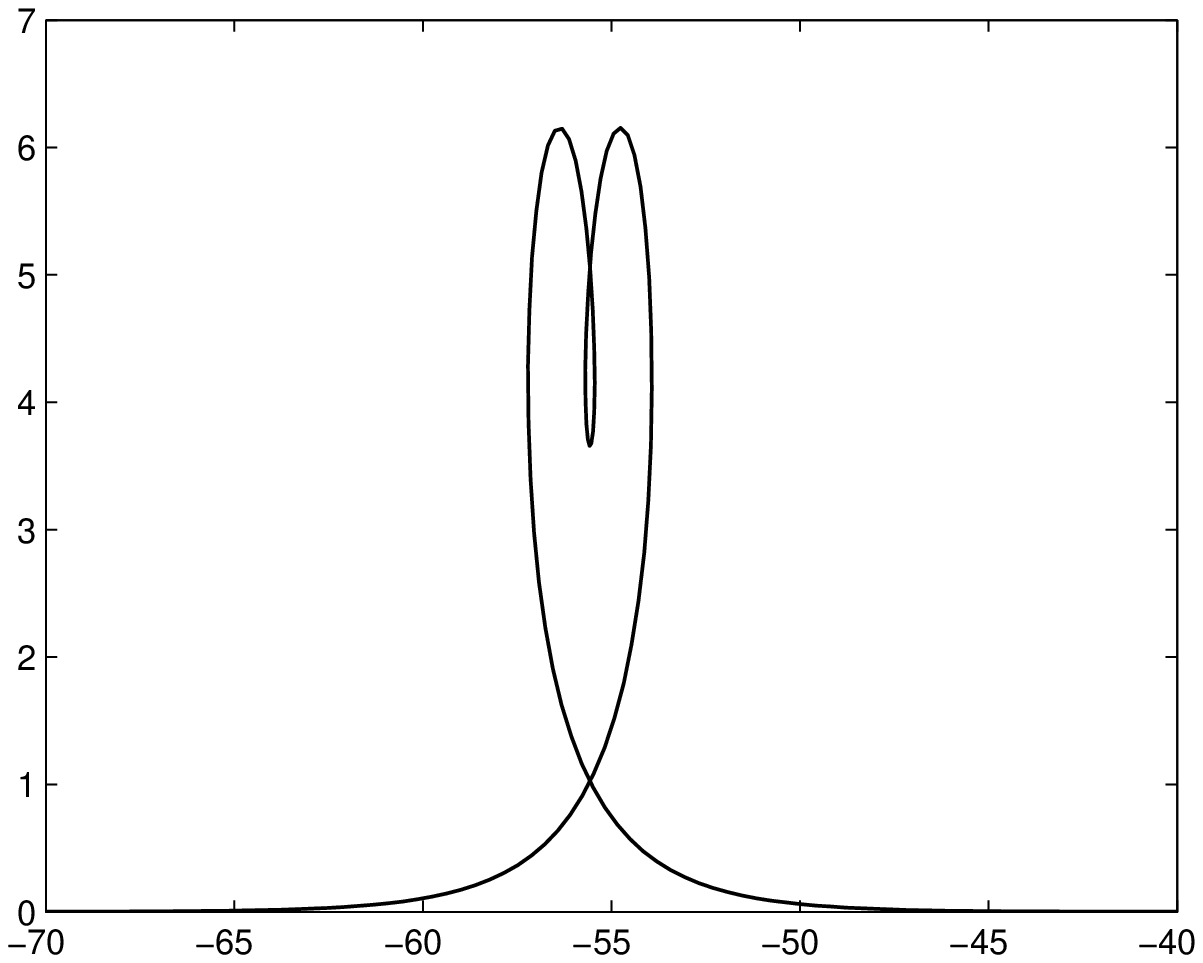}
    \caption{Snapshot of splitting loop-like soliton solution at time $t=-15$ from Eq. (\ref{eq2.1}) for $\alpha=0.1$, $v_{1}=0.24$ and
$v_{2}=0.12$.}\label{FIG6}
\end{center}
\end{figure}

\begin{figure}
    \begin{center}
    \includegraphics{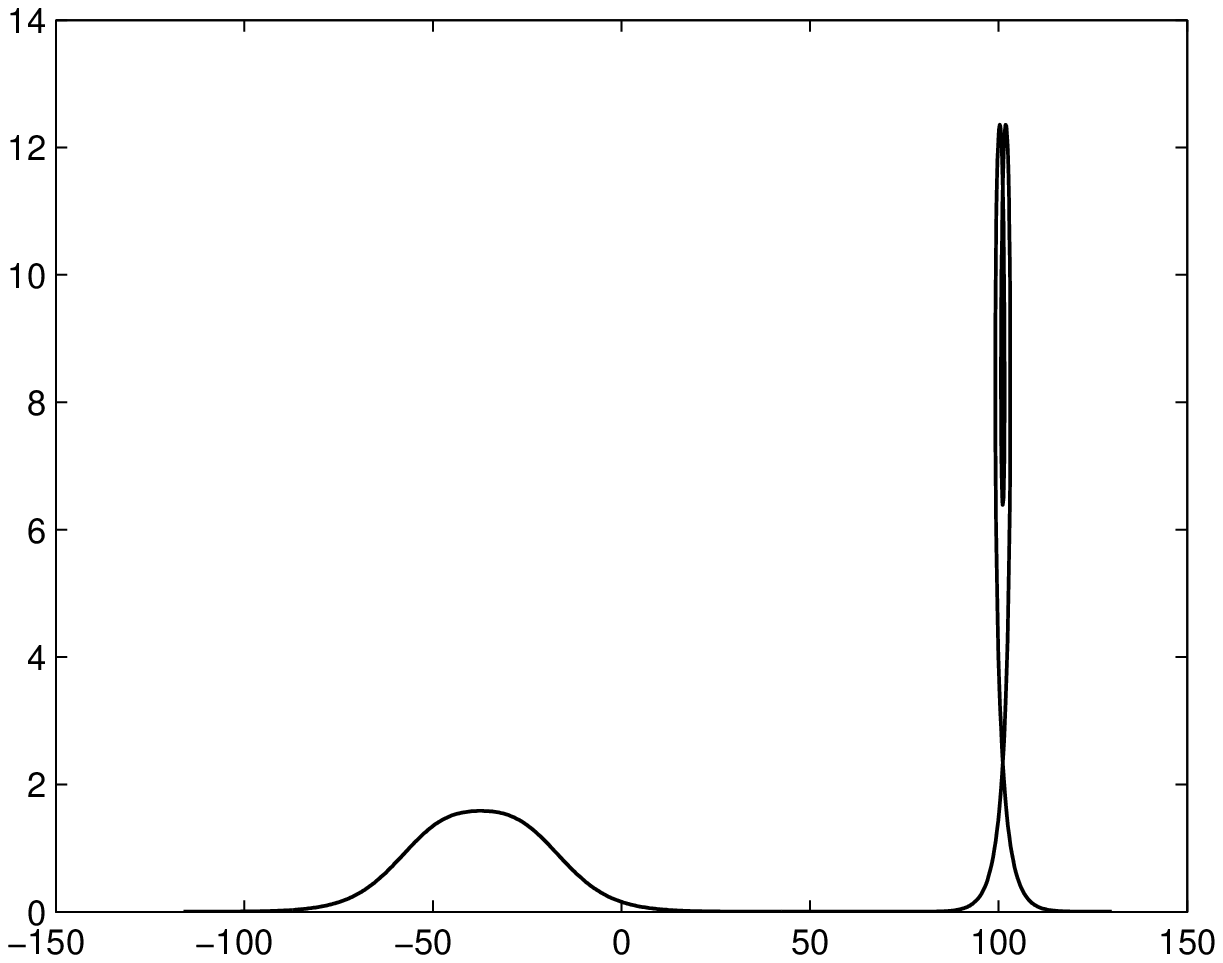}
    \caption{Snapshot of splitting loop-like soliton solution at time $t=11$ from Eq. (\ref{eq2.1}) for $\alpha=0.1$, $v_{1}=0.24$ and
$v_{2}=0.12$.}\label{FIG7}
\end{center}
\end{figure}

\begin{figure}
    \begin{center}
    \includegraphics{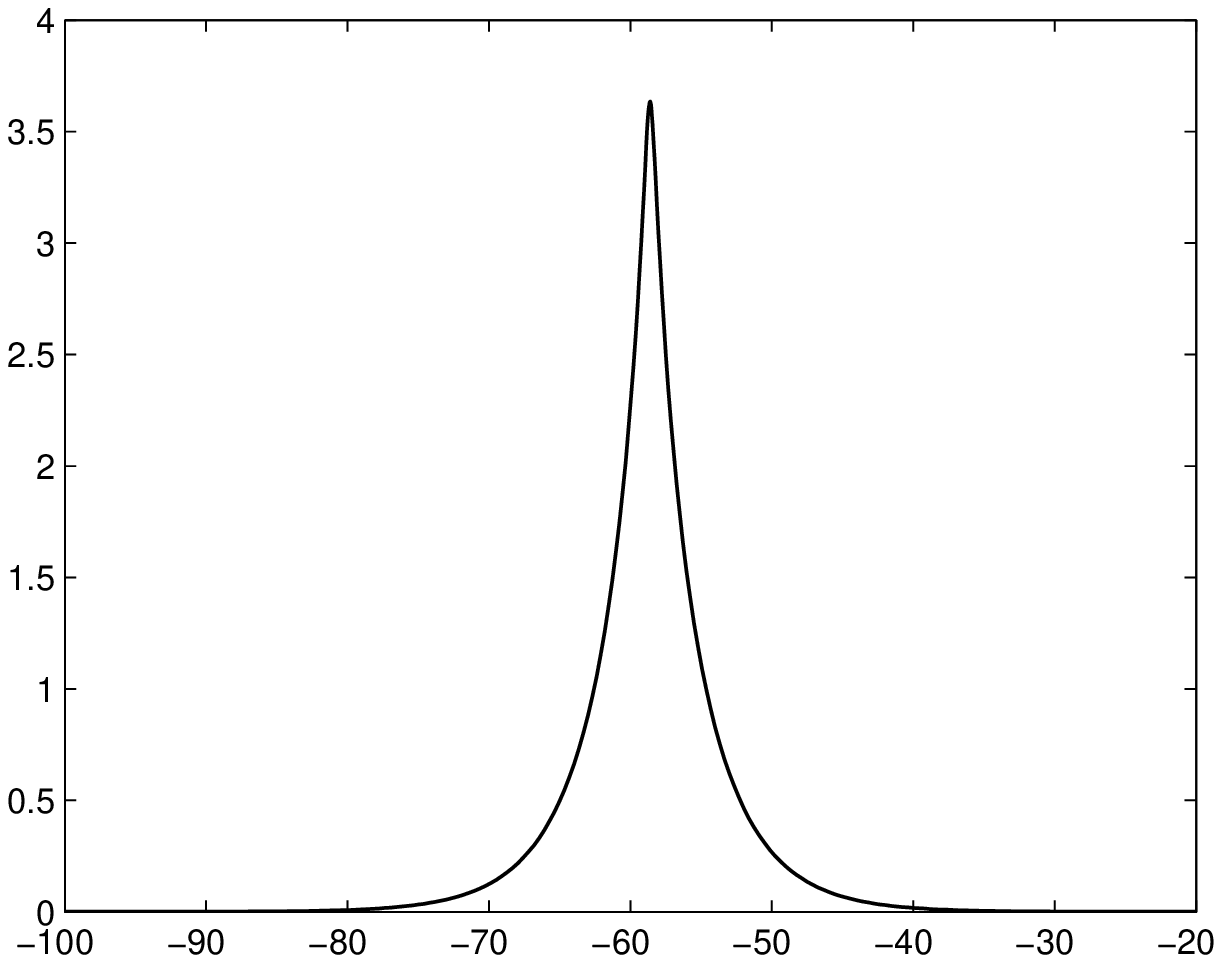}
    \caption{Snapshot of cusp-like soliton solution at time $t=-15$ from Eq. (\ref{eq2.1}) for $\alpha=2.6$, $v_{1}=0.24$ and
$v_{2}=0.12$.}\label{FIG8}
\end{center}
\end{figure}

\begin{figure}
    \begin{center}
    \includegraphics{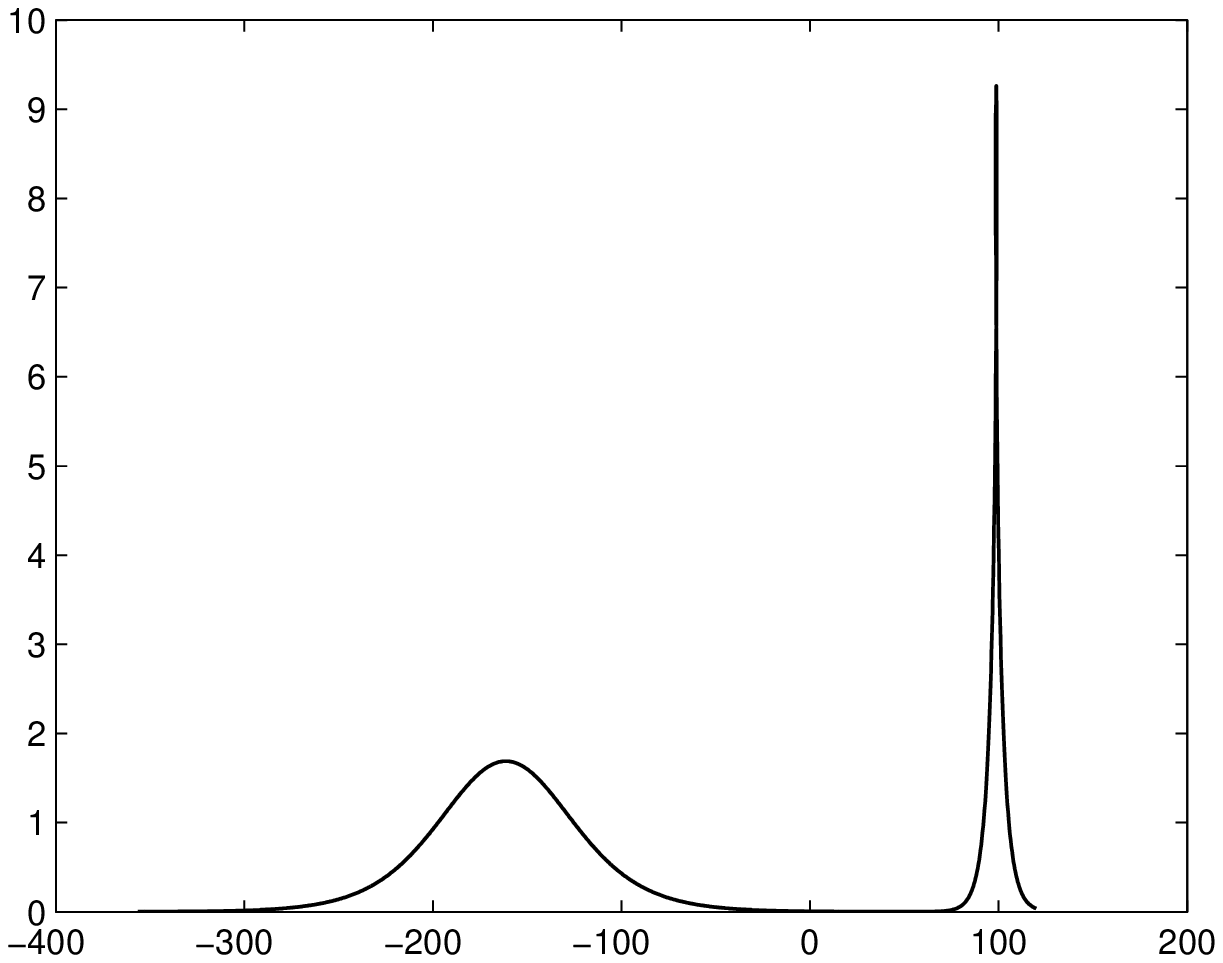}
    \caption{Snapshot of cusp-like soliton solution at time $t=11$ from Eq. (\ref{eq2.1}) for $\alpha=2.6$, $v_{1}=0.24$ and
$v_{2}=0.12$.}\label{FIG9}
\end{center}
\end{figure}

\begin{figure}
    \begin{center}
    \includegraphics{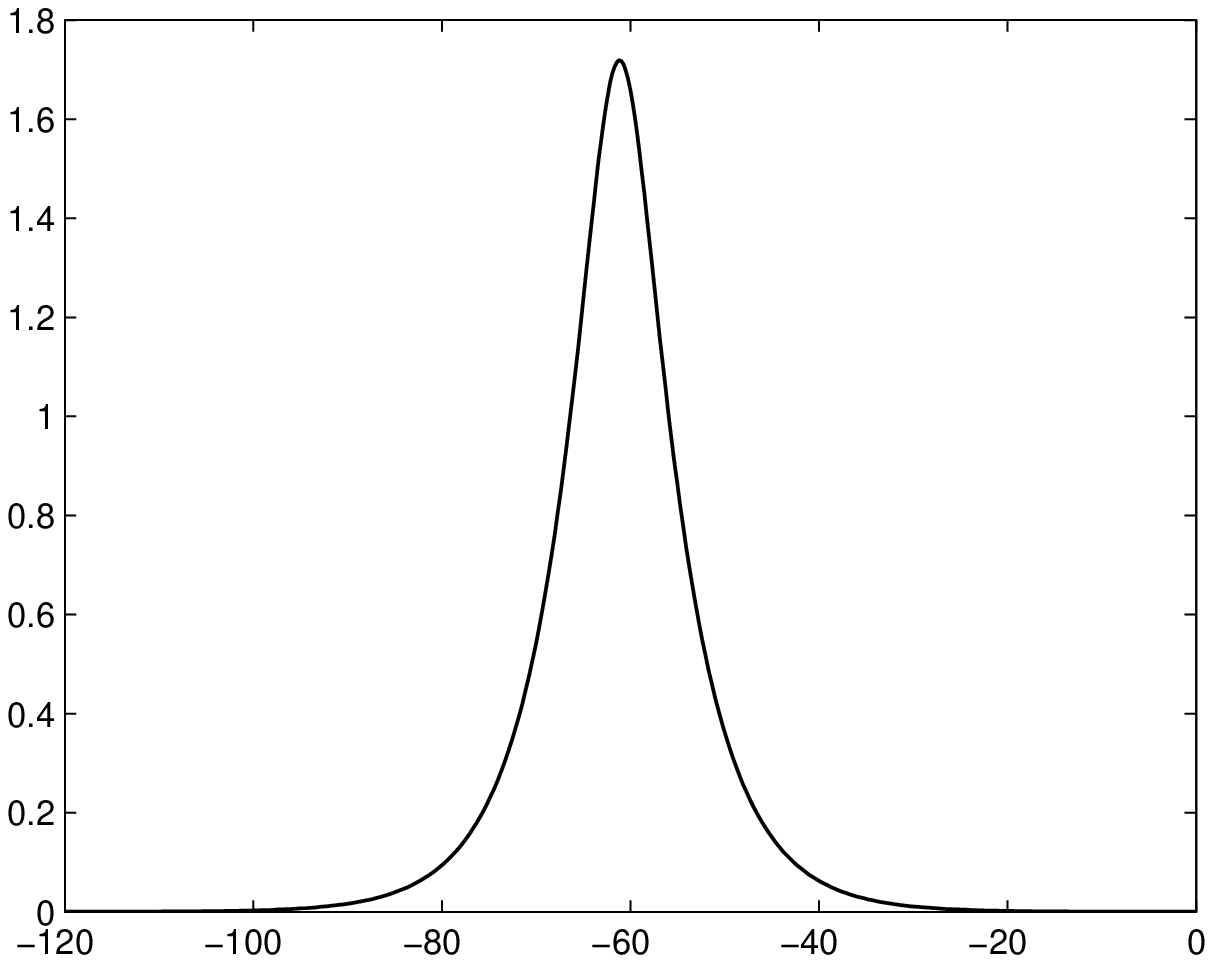}
    \caption{Snapshot of hump-like soliton solution at time $t=-15$ from Eq. (\ref{eq2.1}) for $\alpha=5.0$, $v_{1}=0.24$ and
$v_{2}=0.12$.}\label{FIG10}
\end{center}
\end{figure}

\begin{figure}
    \begin{center}
    \includegraphics{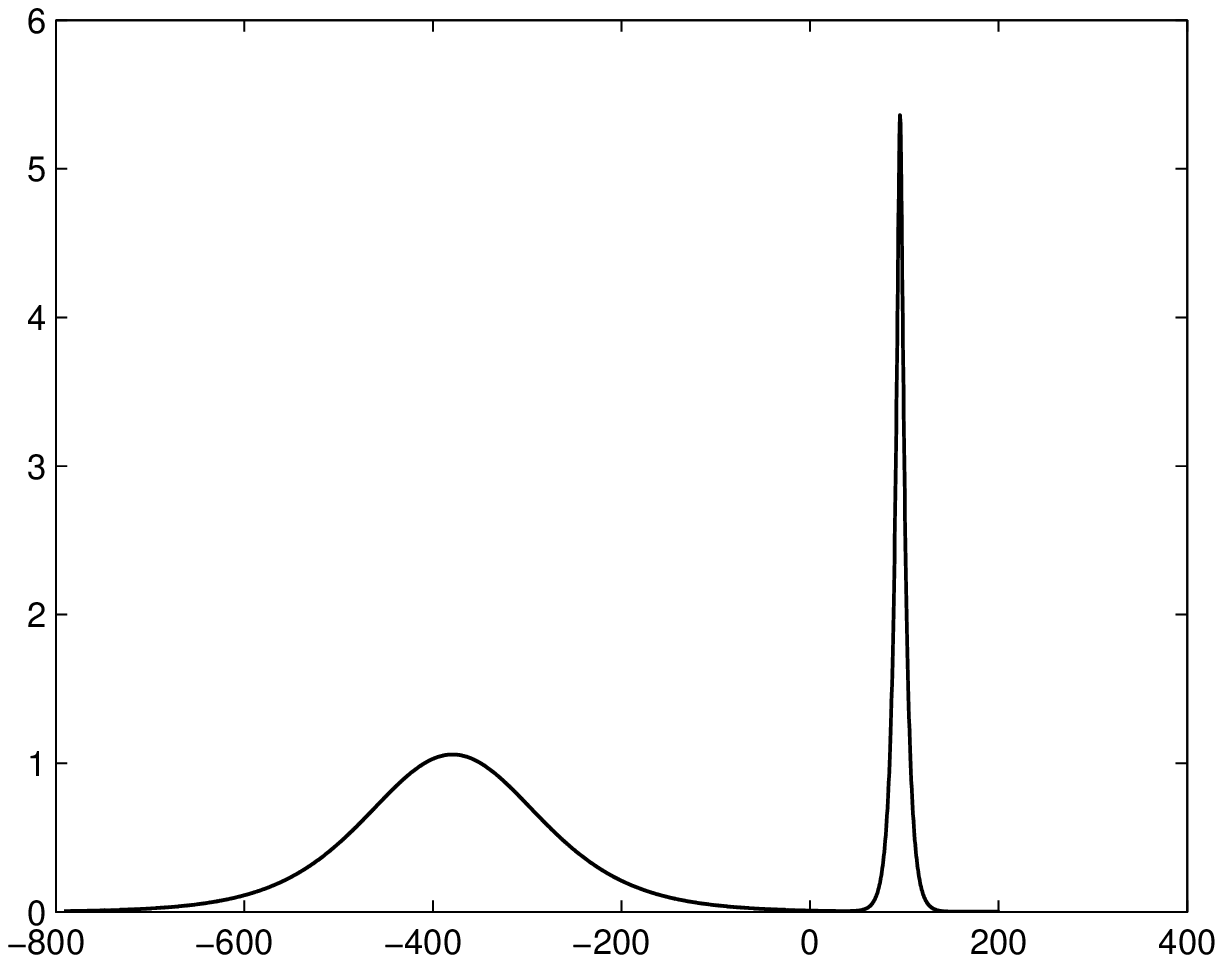}
    \caption{Snapshot of hump-like soliton solution at time $t=11$ from Eq. (\ref{eq2.1}) for $\alpha=5.0$, $v_{1}=0.24$ and
$v_{2}=0.12$.}\label{FIG11}
\end{center}
\end{figure}

%
%
%
%

\end{document}